\documentclass[%
 prl,
 amsmath,amssymb,
 reprint,%
]{revtex4-1}

\usepackage{graphicx}

\usepackage[utf8]{inputenc}
\usepackage[pdfstartview=FitH, pdfpagemode=None, pdfpagelayout=OneColumn, colorlinks=true, linkcolor=blue, citecolor = blue]{hyperref}

\makeatletter
\def\@email#1#2{%
 \endgroup
 \patchcmd{\titleblock@produce}
  {\frontmatter@RRAPformat}
  {\frontmatter@RRAPformat{\produce@RRAP{*#1\href{mailto:#2}{#2}}}\frontmatter@RRAPformat}
  {}{}
}%
\makeatother

\begin{document}

\title{Transformer ratio growth due to ion motion in plasma wakefield accelerators}
\author{V.A. Minakov}%

\author{K.V. Lotov}%
\affiliation{ 
Novosibirsk State University, Novosibirsk, 630090, Russia
}%

\date{\today}

\begin{abstract}
We report a recently discovered counterintuitive effect where breaking of a Langmuir wave in a plasma wakefield accelerator leads to an increase in the accelerating field rather than wave dissipation. 
The effect relies on the ability of transversely breaking waves to draw wave energy from nearby regions due to the inflow of electrons oscillating collectively and the outflow of electrons moving individually.
\end{abstract}

\maketitle

Wakefield acceleration of particles in plasmas is now considered and studied as a possible way to future very high energy accelerators \cite{NJP23-031101, PoP27-070602}.
However, the plasma wave that accelerates the particles also decelerates the drive beam that creates the wave. 
For electron drivers, this imposes a constraint known as the transformer ratio limit, according to which the energy gain of the accelerated (witness) beam cannot be much higher than the driver energy. 
The limitation arises because of the ratio of electric fields, accelerating to decelerating, which for longitudinally symmetric drivers cannot exceed 2 \cite{PAcc17-171}.
Increasing the transformer ratio to a value greater than 2 \cite{PRA33-2056,PoP12-053105,PRST-AB18-081301,NIMA-432-202,PoP22-123113} 
requires precise control of the driver shape \cite{NIMA-829-17,NIMA-865-139,PRL118-104801,NIMA-909-107,PTRSA377-20180392}
and is difficult in practice \cite{IPAC18-1648}, 
although feasible \cite{PRST-AB13-111301,PRL121-064801, PRL124-044802}.
Therefore, even a relatively small increase in the transformer ratio is welcome, as allows a proportional increase in the witness energy at a fixed driver energy.

Recently it was noticed \cite{PPCF61-114003} that the motion of plasma ions can boost the longitudinal electric field by up to 40\%, thus increasing the transformer ratio at no cost, but the reasons for the field growth have remained unclear.
In this paper we show that the field growth is not a direct consequence of ion motion, but a result of local wavebreaking caused by inhomogeneities of ion density. The wavebreaking causes convective transfer of the wave energy towards the axis which, in some cases, can overcome dissipation of the wave energy.
The study is based on simulations with the axisymmetric quasistatic code LCODE \cite{PRST-AB6-061301,NIMA-829-350}.

\begin{figure}[b]
 \includegraphics{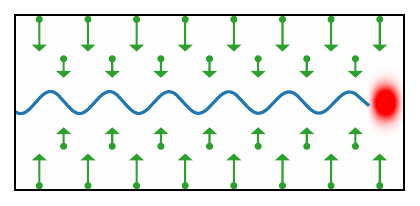}
\caption{Illustration of the simplified model showing the driver in red, the wakefield in blue, and the plasma particles and their initial momentum in green.}
\label{fig1-toy}
\end{figure}
\begin{figure}[b]
\includegraphics{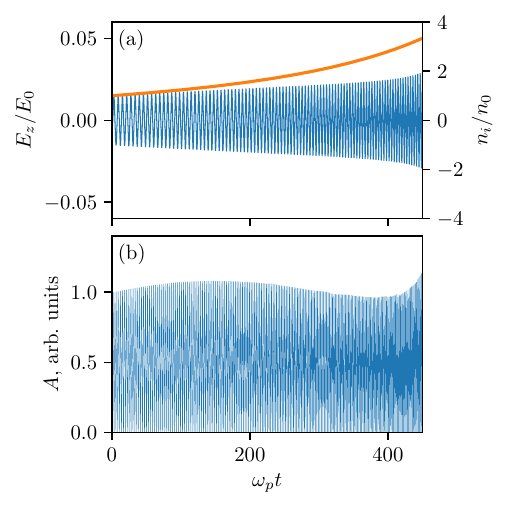}
\caption{The simplified model: time dependences of (a) electric field $E_z$ (blue line), ion density $n_i$ (orange line), and (b) ratio of field energy density to ion density $A = E_z^2/(8\pi n_i)$ on the axis. Time is counted from the moment when the driver center passes.}
\label{fig2-effect}
\end{figure}

First, let us show that convective motion of plasma electrons as a whole can result in increase of the wave amplitude, if some wave was excited in the plasma. Consider the following simplified model (Fig.\,\ref{fig1-toy}). 
Let us impart an inward-directed velocity to plasma ions and electrons, which linearly depends on the radius $r$:
\begin{equation}\label{e1}
    v_r = - 0.001 r \omega_p,
\end{equation}
where $\omega_p$ is the plasma frequency corresponding to the initial plasma density $n_0$.
Then excite a wakefield in this compressing plasma with a short positively charged bunch.
The bunch propagates in the positive $z$ direction without changing shape, and its density is
\begin{equation}\label{e2}
    n_b = 0.01 n_0 e^{-r^2/2 \sigma_r^2} \left[  1 + \cos \left( \sqrt{\frac{\pi}{2}} \frac{\xi}{\sigma_z}  \right)  \right]
\end{equation}
at $|\xi| < \sigma_z \sqrt{2\pi}$ and zero otherwise. Here $\sigma_r = \sigma_z = c/\omega_p$, $c$ is the speed of light, and $\xi = z - ct$ is the co-moving coordinate. 
In a stationary plasma of density $n_0$ this bunch would excite a linear plasma wave with an approximately constant amplitude of the longitudinal electric field of about $0.015E_0$, where $E_0 = mc\omega_p/e$ is the wavebreaking field, $m$ is the electron mass, and $e$ is the elementary charge \cite{NIMA-909-446}.
In the compressing plasma, the field amplitude grows as the plasma density increases [Fig.\,\ref{fig2-effect}(a)], and the ratio of field energy density to plasma density remains approximately constant [Fig.\,\ref{fig2-effect}(b)].
From this observation we conclude that the wave energy is convectively transported by the moving plasma, and the field amplitude can increase if the plasma motion is not divergence-free.

\begin{table}[tb]
\begin{center}
\caption{Simulation parameters}\label{t1}
\begin{tabular}{lll}\hline
  Parameter & \quad & Value \\ \hline
  Plasma ion-to-electron mass ratio && 157\,000\\
  Maximum initial beam density && $0.01 n_0$ \\
  Initial beam half-length && $350 c/\omega_p$  \\
  Initial beam radius && $0.75 c/\omega_p$  \\
  Initial beam energy && 400\,GeV  \\
  Initial angular spread of the beam && $3.7 \times 10^{-5}$  \\
  Location of observation point, $z_0$ && $25\,000c/\omega_p$  \\
  \hline
 \end{tabular}\end{center}
\end{table}

Next, we revisit the case described in Ref.~\onlinecite{PPCF61-114003}, where accumulation of the wave energy near the axis and wakefield growth were first observed. 
Here, a long half-cut proton beam with initial parameters close to those of the AWAKE experiment \cite{Nat.561-363,PPCF62-125023,Symmetry14-1680} (Table~\ref{t1}) self-modulates in the plasma \cite{PRL104-255003,PoP22-103110} and excites a wakefield with a nonzero average electric field, which cause the ions to move \cite{PRL109-145005,PoP21-056705,PPCF64-045003}. 

In the quasistatic model used, the plasma response is calculated in coordinates $(\xi, r)$ periodically along the coordinate $z$. 
Simulation grid step is $0.005 c/\omega_p$ for both $r$- and $\xi$-coordinates. 
There are 10 radially weighted macroparticles of each sort (electrons and ions) per cell.
The plasma and beam states are updated every $200 c/\omega_p$ along the $z$ coordinate.
The simulation window has a width of $20 c/\omega_p$, which is necessary to correctly account for the electron halo surrounding the plasma \cite{PPCF63-055002}.
Although the code outputs plasma characteristics as functions of $\xi = z-ct$, we present the simulation results as functions of time by assuming the coordinate $z$ to be fixed.


\begin{figure}[tb]\centering
 \includegraphics{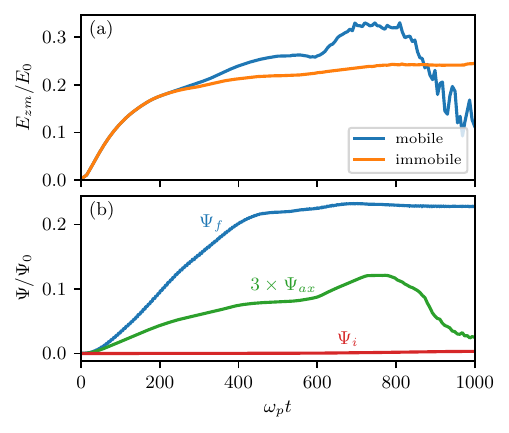}
\caption{(a) The electric field amplitude on the axis ($E_{zm}$) for mobile and immobile ions at $z = z_0$; (b) the corresponding energy fluxes: total $\Psi_f$, in the near-axis region $\Psi_\text{ax}$ (multiplied by 3 for better visibility), and total kinetic energy of ion motion $\Psi_i$. Time is measured from passage of the beam leading edge.}
\label{fig3-awake}
\end{figure}

The effect of interest appears at full beam microbunching and is best seen at the distance $z_0 = 25000 c/\omega_p$ from the entrance to the plasma.
The wakefield in this cross-section first grows being excited by the microbunches, then increases further (at $t \gtrapprox 600 \omega_p^{-1}$) for reasons we have to clarify, after which rapidly decreases (at $t \gtrapprox 800 \omega_p^{-1}$) [Fig.\,\ref{fig3-awake}(a)]. 
The latter two effects are obviously related to ion motion, since they disappear for immobile ions.

To understand the effects, we need to know how the wave energy is distributed in space and time. 
For this purpose, we consider the integral energy fluxes in the co-moving simulation window \cite{PRE69-046405,PoP25-103103,PRL130-105001}:
\begin{equation}
    \Psi (\xi) = \int \left( \frac{E^2 + B^2}{8\pi} - \frac{1}{4\pi} \left[ \vec{E} \times \vec{B} \right]_z \right) c \, dS + \Psi_e + \Psi_i,
\end{equation}
\begin{equation}
    \Psi_\alpha (\xi) = \sum_j (\gamma_j - 1) m_j c^2,
\end{equation}
where $\vec{E}$ and $\vec{B}$ are electric and magnetic fields, $m_j$ and $\gamma_j$ are mass and relativistic factor of plasma particles of the sort $\alpha = e,i$, the integration is performed over a transverse cross-section located at some $\xi$, and the summation is over all plasma particles which intersect this cross section ($\xi = \text{const}$) in unit time. 
The natural unit for the energy flux is $\Psi_0 = m^2 c^5 / (4 \pi e^2)$.
The quantity $\Psi/c$ represents the linear energy density left in the plasma by the beam.

The energy flux $\Psi_f$ integrated over the full plasma cross-section grows when the beam drives the wave and then remains constant [Fig.\,\ref{fig3-awake}(b)]. 
It practically does not change at $t > 600 \omega_p^{-1}$, indicating that at this time the field growth is not caused the beam.
The energy flux $\Psi_\text{ax}$ through the circle $r < c/\omega_p$ however increases after $t \approx 600 \omega_p^{-1}$, indicating energy transfer to the near-axis area in agreement with the observations reported in Ref.~\onlinecite{PPCF61-114003}.
The energy flux $\Psi_i$ associated with the ion motion is negligibly small compared to above two, so the transferred energy is not the kinetic energy of ions.

\begin{figure}[!]\centering
 \includegraphics{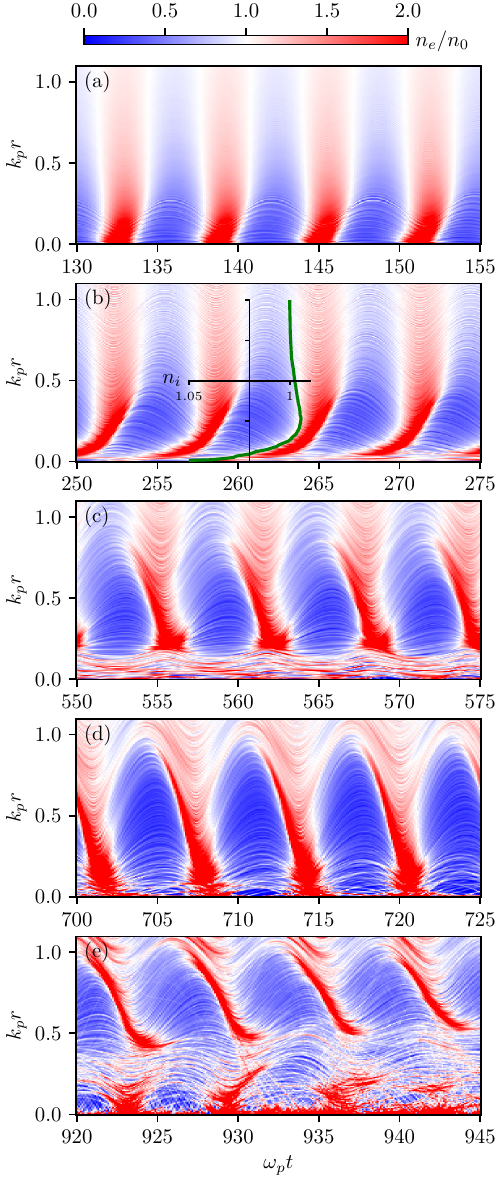}
\caption{Stages of plasma wave evolution on the electron density map at $z=z_0$: (a) regular wave, (b) wavefronts are distorted due to increased ion density near the axis, (c) wavebreaking, (d) wavefronts are restored, (e) ultimate wavebreaking. The green line with small axes superimposed on fragment (b) shows the radial profile of the ion density $n_i$ at $t=260\omega_p^{-1}$.} 
\label{Fig4-ne}
\end{figure}

The process underlying the redistribution of wave energy is best seen in maps of the plasma electron density (Fig.\,\ref{Fig4-ne}). 
In the stage of fast initial growth, the wave is regular and smooth [Fig.\,\ref{Fig4-ne}(a)]. 
A non-zero average radial force gradually perturbs the ion background, and soon an ion density peak appears near the axis [inset in Fig.\,\ref{Fig4-ne}(b)]. 
The increased ion density leads to a shortening of the plasma wavelength near the axis and a corresponding distortion of the wave fronts [Fig.\,\ref{Fig4-ne}(b)]. 
When the curvature of the wave fronts reaches some level, the wave breaks transversely \cite{PRL78-4205,PoP29-023104}, and clearly visible density crests and valleys disappear in the wavebreaking region [Fig.\,\ref{Fig4-ne}(c)]. 
After some time, however, the wave recovers and returns to its original appearance [Fig.\,\ref{Fig4-ne}(d)]. 
Cycles of local wave breaking and recovery can repeat several times (not shown in the figure). 
Each cycle results in a loss of wave energy, which is taken away by the electrons involved in the wavebreaking. 
Finally, the wave weakens and no longer produces a strong electric field [Fig.\,\ref{fig3-awake}(a)], but it still has regions of broken and recovered wave [Fig.\,\ref{Fig4-ne}(e)]. 
The main conclusion from the above observations is that the transverse wavebreaking does not cause the wave to disappear. 
The wave can recover by losing some energy. 
The recovery does not require energy replenishment from the drive beam and is possible behind the driver.  

\begin{figure}[tb]\centering
 \includegraphics{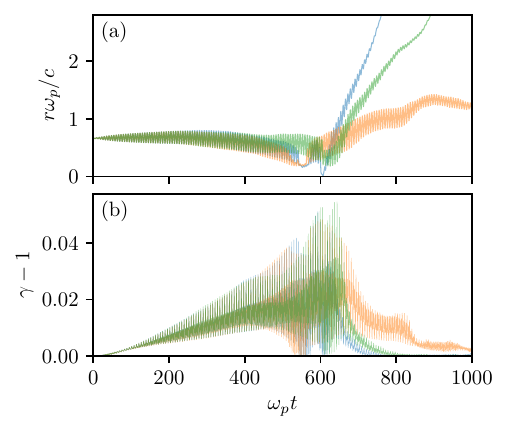}
\caption{(a) The trajectories of three neighboring plasma electrons in the simulation window and (b) the energies of these electrons expressed in terms of their relativistic factor $\gamma$.}
\label{Fig5-trajectories}
\end{figure}

To trace how the wavebreaking can cause the field to grow, let us consider the motion of individual plasma electrons (Fig.\,\ref{Fig5-trajectories}).
Electrons that have fallen out of collective motion as a result of wavebreaking leave the near-axis region.
For the considered electrons, this occurs at $\omega_p t > 600$ [Fig.\,\ref{Fig5-trajectories}(a)].
When some of the electrons leave the wave, the imbalanced ion charge attracts electrons from larger radii toward the axis.
Our electrons plays this role (are attracted) at $250 < \omega_p t < 550$, when the wavebreaking affects the inner electrons.
The electrons attracted towards the axis bring the plasma wave with them.
Until they reach the wavebreaking region, they oscillate coherently and with an increasing energy, which indicate an increase in the wave amplitude at their location [Fig.\,\ref{Fig5-trajectories}(b)].
After the wavebreaking, their trajectories diverge and the electrons leave the wave individually, carrying significantly less energy than they brought to the axis being a constituent of the wave.
Thus, the region of transverse wavebreaking acts as a ``pump'', drawing in electrons and wave to the axis and expelling electrons without the wave.

To summarize, transverse breaking of a plasma wave can initiate pumping of the wave energy to the axis. 
The resulting energy flux can exceed the inevitable energy dissipation that accompanies wavebreaking, leading to an increase in the longitudinal electric field on the axis.

This work was supported by the Russian Science Foundation, Project No. 23-12-00028. Simulations were performed on HPC cluster ``Akademik V.\,M.\,Matrosov'' \cite{Matrosov}.

\end{document}